\documentclass[
twocolumn,
superscriptaddress,
amsmath,amssymb,
prl,
]{revtex4-2}
\usepackage{graphicx}
\usepackage{hyperref}
\usepackage{dcolumn}
\usepackage{bm}
\usepackage{epstopdf}
\usepackage{romannum}
\usepackage{csquotes}
\usepackage[dvipsnames]{xcolor}
\usepackage{hyperref}
\usepackage{titlesec}

\newcommand{\titlecase}[1]{%
	\MakeUppercase{#1}\MakeLowercase{\expandafter\lowercase\expandafter{\substring{#1}{2}{9999}}}}

\begin{document}

\preprint{La0.8Sr0.2NiO2}

\title{Upper critical fields in high-$ T_{\rm{c}} $ superconductors}

\author{Wei Wei}
\affiliation{Department of Physics, Southeast University, Nanjing 211189, China}

\author{Yuling Xiang}
\affiliation{Department of Physics, Southeast University, Nanjing 211189, China}

\author{Qiang Hou}
\affiliation{Department of Physics, Southeast University, Nanjing 211189, China}

\author{Yue Sun}

\email{Corresponding author: sunyue@seu.edu.cn}
\affiliation{Department of Physics, Southeast University, Nanjing 211189, China}

\author{Zhixiang Shi}
\email{Corresponding author:zxshi@seu.edu.cn}
\affiliation{Department of Physics, Southeast University, Nanjing 211189, China}

\setcounter{secnumdepth}{2} 
\begin{abstract}
	
\textbf{}
Since the discovery of high-temperature superconductivity in cuprates, understanding the unconventional pairing mechanism has remained one of the most significant challenges. The upper critical field ($H_{\rm{c2}}$) is an essential parameter for obtaining information on the pair-breaking mechanism, coherence length $\xi$, and pairing symmetry, all of which are crucial for understanding unconventional superconducting mechanisms. Here, we provide a brief review of studies on $ H_{\rm{c2}} $ in several representative series of cuprate, iron-based, and nickelate superconductors. By comparing the behavior of $ H_{\rm{c2}} $ as a function of temperature, doping concentration, and anisotropy across these three major classes of superconductors, we hope to contribute to a better understanding of the complex pairing interactions in high-temperature superconductors.
	\\
	
Keywords: high-temperature superconductivity; cuprate superconductors; iron-based superconductors; nickelate superconductors
	\color{black}
\end{abstract}

\maketitle
\textbf{}
\section{Introduction}
In 1986, Bednorz and Müller made the groundbreaking discovery of high-temperature superconductivity in the Ba-doped insulating cuprate system La$_{2-x}$Ba$_x$CuO$_{4}$ \cite{jg1986possibl}. By 2008, the discovery of iron-based superconductors (FeSCs) provided new insights into the pairing mechanisms of high-temperature superconductors (HTS) and fueled the pursuit of materials with higher $T_{{\rm{c}}}$ \cite{kamihara2006iron,kamihara2008iron}. Over approximately three decades of research, cuprates and FeSCs have been recognized as the only two systems of high-temperature superconductors under ambient pressure. Ongoing discussions regarding the superconducting pairing mechanisms have led to the proposal of various superconducting pairing models that do not involve electron-phonon coupling. Until 2019, Li $ et $ $ al. $ discovered superconductivity in Sr-doped NdNiO$ _{2} $ at temperatures between 9 and 15 K \cite{RN1471}. More recently, in 2023, Wang $et$ $al.$ identified signatures of superconductivity near 80 K in La$_3$Ni$_2$O$_7$ under high pressure, suggesting that nickelate superconductors may represent a new class of high-temperature superconductors \cite{80K}. These findings provide a new platform to explore the mechanisms of high-temperature superconductivity, further stimulating research into the pairing mechanisms underlying high-temperature superconductivity. 

Upper critical field, $H_{\rm{c2}}$, is a fundamental parameter for getting information such as the pair-breaking mechanism, coherence length $\xi$, anisotropy, and pairing symmetry — all of which play important roles in understanding unconventional superconductivity \cite{RN1488,zhuang2015pauli,RN1499}. In type-II superconductors, the origins of $H_{\rm{c2}}$ are classified as two distinct pair-breaking mechanisms. 
One is the orbital pair-breaking effect, where the Lorentz force acts on Cooper pairs with opposite momenta, and the superconductivity is destroyed when the kinetic energy exceeds the condensation energy of the Cooper pairs. The other is the spin-paramagnetic pair-breaking effect, which arises from the Zeeman splitting of spin-singlet cooper pairs. In this case, superconductivity is suppressed when the Pauli spin magnetization exceeds the condensation energy as the critical magnetic field is reached. 
The analysis and comparison of $H_{\rm{c2}}$ behaviors in the three major classes of unconventional superconductors—cuprates, FeSCs, and nickelates—are crucial for understanding the mechanisms behind unconventional pairing.

To begin with, it is essential to compare the physical properties among these three types of unconventional HTS. The major comparisons can be summarized as follows: (i) All these superconductors are layered compounds, with crystal structures containing CuO, FeAs/FeSe, or NiO layers. Various atoms or molecules can be intercalated between these layers, altering the lattice constants or even modifying the crystal structures, which leads to the formation of various superconducting phases. (ii) The parent compounds of cuprates and infinite-layer nickelates are Mott insulators \cite{keimer2015quantum,goodge2021doping}, with superconductivity emerging upon chemical doping. In contrast, the parent compounds of FeSCs are antiferromagnetism \cite{ren2009research}. (iii) In terms of electronic structure, x-ray absorption spectroscopy and resonant inelastic x-ray scattering have shown that LaNiO$ _{2} $ and NdNiO$_{ 2 }$ have a nominal 3$ d^{9} $ electronic configuration, analogous to the Cu$ ^{2+} $ state in cuprates \cite{hepting2020electronic}.
However, the Fe$ ^{2+} $ electronic configuration in iron-based superconductors is different, with two unpaired electrons \cite{hu2016identifying}. (iv) Regarding band structure, most cuprates exhibit a single-band characteristic, with intraorbital interactions as the sole driving force for pairing, resulting in a $ d $-wave gap \cite{damascelli2003angle}. Only a few cuprate superconductors, such as Nd$ _{2-x} $Ce$ _{x} $CuO$ _{4+\delta} $, exhibit multi-band characteristics. In contrast, most FeSC and infinite-layer nickelate superconductors are multiband systems, requiring both intraorbital and interorbital interactions for pairing \cite{llovo2021multiband,PhysRevLett.125.027001}. 
\begin{figure*}\center
	\includegraphics[width=1\linewidth]{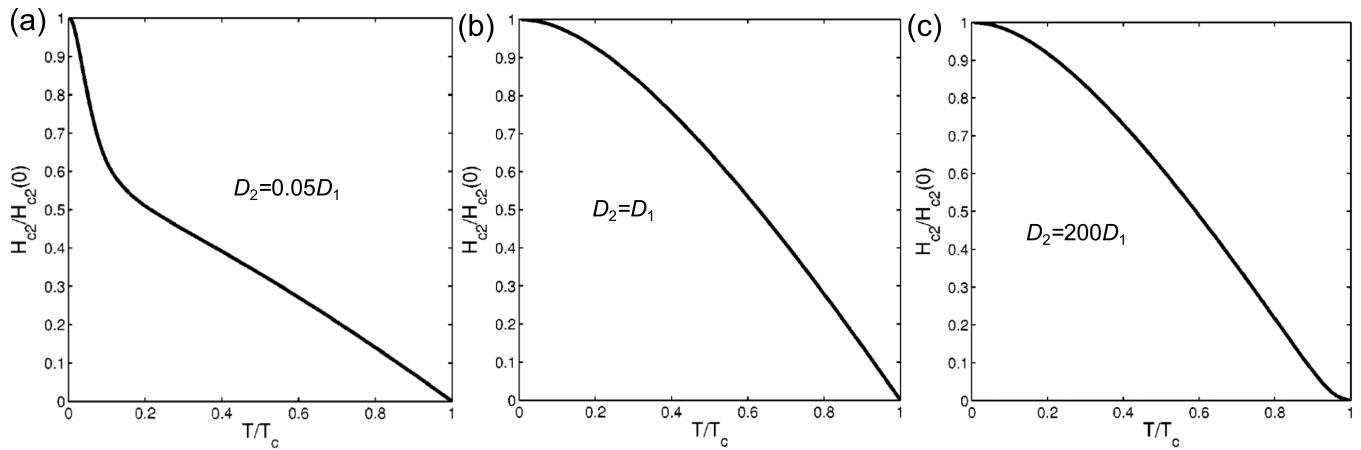}
	\caption{(a) Temperature dependence of $ H_{\rm{c2}}(T) $, calculated from Eq. (10) for various ratios of $ D_2/D_1 $ \cite{PhysRevB.67.184515}, was determined for MgB$_2$ \cite{golubov2002specific}. $ \sigma $ and $ \pi $ represent different orbitals. 	
	}\label{}
\end{figure*}

In this review, we first provide a brief introduction to the theoretical models of $H_{\rm{c2}}$ in Section 2. Subsequently, in Sections 3 to 5, we offer an overview of the temperature and doping dependence, as well as the anisotropy, of $H_{\rm{c2}}$ in cuprate, iron-based, and nickelate superconductors. Finally, we conclude the review with a summary and outlook in Section 6.

\section{Theoretical models of $H_{\rm{c2}}$}
\subsection{Ginzburg-Landau (GL) model}
The Ginzburg-Landau (GL) theory provides a phenomenological framework for understanding superconductivity near $T_{\rm{c}}$. Specifically, within the context of $H_{\rm{c2}}$, the GL theory primarily focuses on orbital effects, which arise from the interaction between Cooper pairs and an applied magnetic field due to their orbital motion. Within this framework, the behavior of $H_{\rm{c2}}$ is closely related to the coherence length $\xi$ and the penetration depth $\lambda$, which characterize the spatial variation of the superconducting order parameter and the magnetic field, respectively. 
\begin{eqnarray}\notag
	\\H_{c2}^{\rm{GL}}(T)=\frac{\phi_{0}}{2\pi\xi^{2}_{ab}(T)},
\end{eqnarray}

where $\Phi_0$ denotes the magnetic flux quantum. 
In a general scenario, an anisotropic superconductor is characterized by direction-dependent $H_{\rm{c2}}$ and two distinct coherence lengths, depending on the orientation of the magnetic field along different high-symmetry axes. First, when the magnetic field is oriented parallel to the crystallographic $ c $ axis ($H \parallel c$), the GL coherence length of Cooper pairs forming screening orbits in the $ ab $-plane can be expressed as \cite{PHD}
\begin{eqnarray}\notag
	\\\xi_{ab}^{\rm{GL}}= \sqrt{\frac{\phi_{0}}{2\pi T_{\rm{c}}\mu_{0}\mid\frac{dH_{\rm{c2}}^{c}}{dT}\mid_{T_{\rm{c}}}}}  .
\end{eqnarray}
Here, $\mu_{0} \frac{dH_{\rm{c2}}^{c}}{dT} $ is the slope of $ H_{\rm{c2}}^{c} $ ($H \parallel c$) in the limit of $ T $ near $T_{\rm{c}}$. For the case where the magnetic field is oriented along the $ ab $-plane ($H \parallel ab$), the coherence length along the $ c $ axis can be expressed as: 
\begin{eqnarray}\notag
	\\\xi_{c}^{\rm{GL}}=\frac{\phi_{0}}{2\pi T_{\rm{c}}\mu_{0}\mid\frac{dH_{\rm{c2}}^{ab}}{dT}\mid_{T_{\rm{c}}}\xi_{ab}^{\rm{GL}}}  .
\end{eqnarray}
In this scenario, $ \mu_{0}\mid\frac{dH_{\rm{c2}}^{ab}}{dT}\mid_{T_{\rm{c}}}$ is the slope of $H_{\rm{c2}}^{ab}$ near $ T_{\rm{c}} $. One issue within the GL model for anisotropic superconductors is the assumption of a linear temperature dependence of $H_{\rm{c2}}$. At low temperatures, this assumption often leads to an overestimation of $H_{\rm{c2}}$, resulting in an underestimation of the coherence lengths. To more accurately determine the coherence lengths, measurements of $H_{\rm{c2}}$ at low temperatures are essential. When measured at $ T \rightarrow 0 $ the equations can be modified as follows:
\begin{eqnarray}\notag
	\\\xi_{ab}^{\rm{GL}}= \sqrt{\frac{\phi_{0}}{2\pi\mu_{0}H_{\rm{c2}}^{c}(T \rightarrow 0)}} 
\end{eqnarray}
and
\begin{eqnarray}\notag
	\\\xi_{c}^{\rm{GL}}=\frac{\phi_{0}}{2\pi\mu_{0}H_{\rm{c2}}^{ab}(T \rightarrow 0)\xi_{ab}^{\rm{GL}}}  .
\end{eqnarray}
Here, $\xi_{ab}$ and $ \xi_{c}$ are the true coherence lengths at zero temperature, with $ H_{\rm{c2}}^{c}(T \rightarrow 0) $ representing the upper critical field as the temperature approaches zero.
However, this approach does not account for factors like Pauli paramagnetism, which may alter the upper critical field.

\subsection{Single-band WHH model}
In 1966, Werthamer, Helfand, and Hohenberg developed a comprehensive model that accounts for the orbital effect, Pauli paramagnetism, and spin-orbit scattering, known as the WHH model for $H_{\rm{c2}}$. This theory is applicable to single-band superconductors in the dirty limit ($ l  \ll \xi $). By incorporating the Maki parameter $ \alpha_{\rm{M}} $, representing the relative strength of orbital and spin pair-breaking, and $ \lambda _{\rm{so}} $, representing spin-orbit scattering, the relationship between temperature and $H_{\rm{c2}}$ can be expressed using the double gamma function \cite{werthamer1966temperature}:
\begin{eqnarray}\notag
	\\\ln(\frac{1}{t})=(\frac{1}{2}+\frac{i\lambda_{\rm{so}}}{4\gamma})\psi(\frac{1}{2}+\frac{\overline{h}+\frac{1}{2}\lambda_{\rm{so}}+i\gamma}{2t})+\notag\\(\frac{1}{2}-\frac{i\lambda_{\rm{so}}}{4\gamma})\psi(\frac{1}{2}+\frac{\overline{h}+\frac{1}{2}\lambda_{\rm{so}}-i\gamma}{2t})-\psi(\frac{1}{2}).
\end{eqnarray}
Here, $ t=T/T_{\rm{c}} $ denotes the reduced temperature, $ \psi $ is the digamma function, 
and $ \gamma= [(\alpha_{\rm{M}}\overline{h})^{2}-(\lambda_{\rm{so}}/2)^{2}] $. 
The reduced magnetic field $ \overline{h} $ is defined as $ \overline{h}=(4/\pi^{2})H_{\rm{c2}}/|-dH_{\rm{c2}}/dt|_{t=1} $, where $ dH_{\rm{c2}}/dt|_{t=1} $ represents the slope of $H_{\rm{c2}}$ near $T_{\rm{c}}$. The above equation can be numerically solved to determine the temperature dependence of $H_{\rm{c2}}$. By fitting the experimental $H_{\rm{c2}}$($ T $) data using WHH theory, the parameters $ \lambda _{\rm{so}} $ and $ \alpha_{\rm{M}} $ can be extracted. In the absence of both the spin-paramagnetic effect and spin-orbit interaction, $ \alpha_{\rm{M}} $ = 0 and $ \lambda _{\rm{so}} $ = 0,  the orbital-limited upper critical field is given by:
\begin{eqnarray}\notag
	\\H_{\rm{c2}}^{\rm{orb}}=-0.693\frac{dH_{\rm{c2}}}{dT}\Big|_{T=T_{\rm{c}}}.
\end{eqnarray}
In the clean limit, the prefactor is replaced by -0.73, meaning that
\begin{eqnarray}\notag
	\\H_{\rm{c2}}^{\rm{orb,clean}}=-0.73\frac{dH_{\rm{c2}}}{dT}\Big|_{T=T_{\rm{c}}}.
\end{eqnarray}
As previously discussed, there are two mechanisms for breaking Cooper pairs in the presence of a magnetic field: the orbital pair-breaking effect and the spin-paramagnetic pair-breaking effect. In the latter case, as the magnetic field exceeds the Zeeman energy, one of the spins in the Cooper pair flips, leading to pair breaking. This critical magnetic field is known as the Pauli-limiting field, and it is determined by the equation \cite{PhysRevLett.9.266,chandrasekhar1962note}:
\begin{eqnarray}\notag
	\\\frac{\chi_{N}(H_{\rm{p}})^{2}}{2}=\frac{N(E_{F})\Delta^{2}}{2},
\end{eqnarray}
where $ \chi_{N}=g\mu_{B}^{2}N(E_{F}) $ represents the spin susceptibility in the normal state. $ g $ is the Lande $ g $ factor, $ \mu_{B} $ is the Bohr magneton, $ \Delta $ is the superconducting gap and $ N(E_{F}) $ is the density of state at the Fermi level $ E_{F} $. In the weak-coupling BCS limit, 2$ \Delta(0)=3.53k_{B}T_{\rm{c}} $, and assuming $ g=2 $, the Pauli-limiting field is given by $ H_{\rm{p}}^{\rm{BCS}}(0)=1.84T_{\rm{c}} $. When considering strong coupling corrections that include electron-boson and electron-electron interactions, the Pauli-limiting field is determined by the expression $ H_{\rm{p}}=1.86(1+\lambda)^{\varepsilon}\eta_{\Delta}\eta_{ib}(1-I) $. The Stoner factor \( I \) is expressed as \( I = N(0)J \), where \( J \) represents an effective exchange integral. The parameter \( \eta_{ib} \) is introduced to describe phenomenologically the effect of gap anisotropy, multiband  character, energy dependence of states etc. The relationship between the orbital limit \(  H_{\rm{orb}}^{c2}(0) \) and the Pauli limit \( H_{\rm{P}}(0) \) is given by: 
\begin{eqnarray}\notag
	\\H^{p}_{c2} (0)= H^{\rm{orb}}_{c2}(0)/(1+\alpha_{\rm{M}}^{2}).
\end{eqnarray}
\subsection{Two-band model}
The aforementioned WHH model applies to single-band superconductors in the dirty limit and has been successfully used to describe the behavior of $ H_{\rm{c2}}(T) $. However, for iron-based superconductors and other systems that exhibit multiband characteristics, the WHH model is not applicable. Instead, a two-band model in the dirty limit, which does not include paramagnetic effects or spin-orbit scattering (unlike the WHH model), has been derived in Ref. \cite{PhysRevB.67.184515}:
\begin{eqnarray}\notag
	\\a_{1}[\ln t+U(h)]+a_{2}[\ln t+U(\eta h)]+\notag\\a_{0}[\ln t+U(h)][\ln t+U(\eta h)]=0,
\end{eqnarray}
where $ t=T/T_{\rm{c}} $, $ h=H_{\rm{c2}}D_{1}/2\Phi_{0}T $, $ a_{0}=2\omega/\lambda_{0} $, $ a_{1}=1+\lambda_{-}/\lambda_{0} $, and $ a_2 = 1 - \lambda_{-}/\lambda_0, \quad \lambda_{\pm} = \lambda_{11} \pm \lambda_{22}, \quad w = \lambda_{11} \lambda_{22} - \lambda_{12} \lambda_{21}$,  $U(x)=\psi(1/2+x)-\psi(1/2) $, and $\quad \lambda_0 = \sqrt{\lambda_{-}^2 + 4 \lambda_{12} \lambda_{21}} $. The function \( \psi(x) \) represents the digamma function, \( \lambda_{11} \) and \( \lambda_{22} \) correspond to the intraband coupling constants, \( \lambda_{12} \) and \( \lambda_{21} \) represent the interband coupling constants, and $ \Phi_{0} $ is the magnetic flux quantum. $ \eta = D_2 / D_1 $, $ \quad D_i $ is the diffusivity of band $ i $.
When the diffusivities are equal ($\eta = 1$), Eq. (10) simplifies to $\ln t + U(h) = 0$, which describes $ H_{\rm{c2}} $ for one-gap dirty superconductors, equivalent to the WHH model in the dirty limit without paramagnetic effects or spin-orbit scattering.
The shape of the $ H_{\rm{c2}}(T) $ curve primarily depends on the ratio of the intraband electron diffusivities $ D_{1} $ (electron diffusivities) and $ D_{2} $ (hole diffusivities), as shown in Fig. 1. The evolution of the temperature dependence of $ H_{\rm{c2}}(T) $ is evident, transitioning from one-band dirty-limit behavior at $ D_1 = D_2 $ to distinct $ H_{\rm{c2}}(T) $ curves. For $ D_1 \gg D_2 $, the curve exhibits a concave shape near $ T_{\rm{c}}$, whereas for $ D_1 \ll D_2 $, an upward curvature is observed at lower temperatures. 

When interband coupling is neglected (\( \lambda_{12} \)=0), Eq. (10) separates into two independent WHH equations \cite{charikova2013upper,RN1488,li2019hole}. For the first type of carriers with diffusion coefficient $ D=D_{1} $, the equation becomes:
\[
\ln(t^1) + U(\eta) = 0,
\]
where $ t^1 = T$/$T_{\rm{c}}^{1} $, $T_c^1 \sim \exp (-1/\lambda_{11})$. For the second type, with \(D = D_2\), Eq. (10) yields: \[
\ln({t^1}) + U(\eta h) = \lambda_0/\omega,
\]
or
\[
\ln({t^2}) + U(\eta h) = 0,
\]
where $ t^2 = T$/$T_{\rm{c}}^{2} $, $T_{\rm{c}}^2 \sim \exp (-1/\lambda_{22})$, $ \lambda_{0}/\omega=(\lambda_{11}-\lambda_{22})/(\lambda_{11}\lambda_{22}) $ and $ T_{\rm{c2}}/T_{\rm{c1}}=\exp (-(\lambda_{11}-\lambda_{22})/(\lambda_{11}\lambda_{22})) $, are used \cite{charikova2013upper}. The two-band fitting has been clearly demonstrated in the cuprate high-temperature superconductor Nd$ _{2-x} $Ce$ _{x} $CuO$ _{4+\delta} $, breaking into two independent WHH components, as shown in the following part of Fig. 3(b) \cite{charikova2013upper}.

To consider the paramagnetic effect and interband scattering, the dirty-limit two-band WHH model is applied \cite{RN1466,RN1420}:
\begin{eqnarray}\notag
	\\(\lambda_{0}+\lambda_{i})[\ln t+U_{+}]+(\lambda_{0}-\lambda_{i})[\ln t+U_{-}]+\notag\\2w[\ln t+U_{+}][\ln t+U_{-}]=0,
\end{eqnarray}
where $ \lambda_{\textpm}=\lambda_{11}\textpm\lambda_{22}, w=\lambda_{11}\lambda_{22}-\lambda_{12}\lambda_{21}, \lambda_{0}=(\lambda_{-}^{2}+4\lambda_{12}\lambda_{21})^{1/2}, \gamma_{\pm}=\gamma_{12}\pm\gamma_{21}, \lambda_{i}=[(w_{-}+\gamma_{-})\lambda_{-}-2\lambda_{12}\gamma_{21}-2\lambda_{21}\gamma_{12}]/\Omega_{0}, 2\Omega_{\pm}=w_{+}+\gamma_{+}\pm \Omega_{0}, \Omega_{0}=[(w_{-}+\gamma_{-})^2+4\gamma_{12}\gamma_{21}]^{1/2}, w\pm=[(D_{1}\pm D_{2})\pi\hslash H_{\rm{c2}}]/\phi_{0}, U\pm=Re\psi(1/2+(\Omega\pm+i\mu_{eff}H_{\rm{c2}})/2\pi k_{B}T)-\psi(1/2) $. Here $ \gamma_{mm^{'}} $ are the inter-band scattering rates and $ \mu_{eff} $ is the electronic effective magnetic moment. 

Finally, considering both orbital and paramagnetic de-pairing effects, the clean-limit two-band WHH model is examined for completeness \cite{PhysRevB.82.184504}:
\begin{eqnarray}\notag
	\\(\ln t + U_{1}(h))(\ln t + U_{2}(\eta h)) + a_{1}(\ln t + U_1(h)) +\notag \\a_{2}(\ln t + U_{2}(\eta h)) = 0	
\end{eqnarray}
\begin{eqnarray*}
	\\	U_{1} = 2 e^{ q^{2}}Re\Sigma^{\infty}_{0} \int^{\infty}_{q}du
	e^{-u^{2}}( \frac{u}{n + \frac{1}{2}} -\\ \frac{t}{\sqrt{b}} \tan^{-1}[\frac{u \sqrt{b}}{t(n + \frac{1}{2}) + i \alpha_{1} b}]),\\
	U_{2} = 2 e^{ q^{2}s}Re\Sigma^{\infty}_{0} \int^{\infty}_{q\sqrt{s}}du
	e^{-u^{2}}( \frac{u}{n + \frac{1}{2}} -\\ \frac{t}{\sqrt{b\eta}} \tan^{-1}[\frac{u \sqrt{b\eta}}{t(n + \frac{1}{2}) + i \alpha_{2} b}]),
\end{eqnarray*}
\begin{eqnarray*}
	\\b=\frac{\hslash ^{2}v_{1}^{2}H_{\rm{c2}}}{8\pi\phi_{0}k_{B}^{2}T_{\rm{c}}^{2}}, \alpha_{i}=\frac{4\mu_{i}\phi_{0}k_{B}T_{\rm{c}}}{\hslash^{2}v_{1}^{2}}, q^{2} = \frac{Q^{2}\phi_{0}\varepsilon_{1}}{2\pi H_{\rm{c2}}}, 
\end{eqnarray*}

\begin{eqnarray*}
	\eta=\frac{v_{2}^{2}}{v_{1}^{2}}, S=\frac{\varepsilon_{2}}{\varepsilon_{1}}, 
\end{eqnarray*}
where $ a_{1}=(\lambda_{0}+\lambda_{-})/2w, a_{2}=(\lambda_{0}-\lambda_{-})/2w,\lambda_{\pm}=\lambda_{11}\pm\lambda_{22}, w=_{11}\lambda_{22}-\lambda_{12}\lambda_{21}$, and $\lambda_{0}=\sqrt{\lambda_{-}^{2}+4\lambda_{12}\lambda_{21}}. $ Here, $ v_{i} $ represents the in-plane Fermi velocity for band $ i $, $ \varepsilon_{i} $ denotes the mass anisotropy ratio ($ \varepsilon_{i}=m_{i}^{\bot}/m_{i}^{\parallel}  $), $ \mu_{i} $ is the effective magnetic moment of the electron, $ Q $ stands for the Fulde–Ferrell–Larkin–Ovchinnikov (FFLO) wave vector.

\subsection{Anisotropy of $ H_{\rm{c2}} $}
The study of the angle dependence of $H_{\rm{c2}}$ is important, as it provides crucial insights into the anisotropy of superconductivity and the superconducting order parameter \cite{gorkov1984anisotropy}. For three-dimensional (3D) bulk superconductors, the angle dependence of $H_{\rm{c2}}$ can be described by the GL model using the following formula \cite{PhysRevLett.68.875,PhysRevB.107.L220503}: 
\begin{eqnarray}\notag
	\\H_{c2}(\theta)=H_{c2}^{ab} / \sqrt{\cos^{2}\theta+(H_{c2}^{ab} / H_{c2}^{c})^{2}\sin^{2}\theta)}.
\end{eqnarray}
The polar angle $ \theta $ represents the angle between the magnetic field and the $ ab $-plane. The Tinkham model describes the behavior of $H_{\rm{c2}}(\theta)$ in two-dimensional systems, which satisfy the condition that the out-of-plane coherence length is larger than the thickness of the superconducting layer ($ d_{\rm{SC}} $):
\begin{eqnarray}\notag
	\\\mid\dfrac{H_{c2}(\theta)\sin\theta}{H_{c2}^{c}}\mid+(\dfrac{H_{c2}(\theta)\cos\theta}{H_{c2}^{ab}})^{2}=1.
\end{eqnarray}
This model accounts for the cusplike behavior observed near parallel magnetic fields, a characteristic commonly seen in various 2D systems \cite{PhysRevB.96.020504,PhysRevB.107.L220503}. It is important to note that the 2D Tinkham model assumes that $H_{\rm{c2}}(\theta)$ is entirely governed by orbital effects across the entire angular range.
\begin{figure}\center
	\includegraphics[width=1\linewidth]{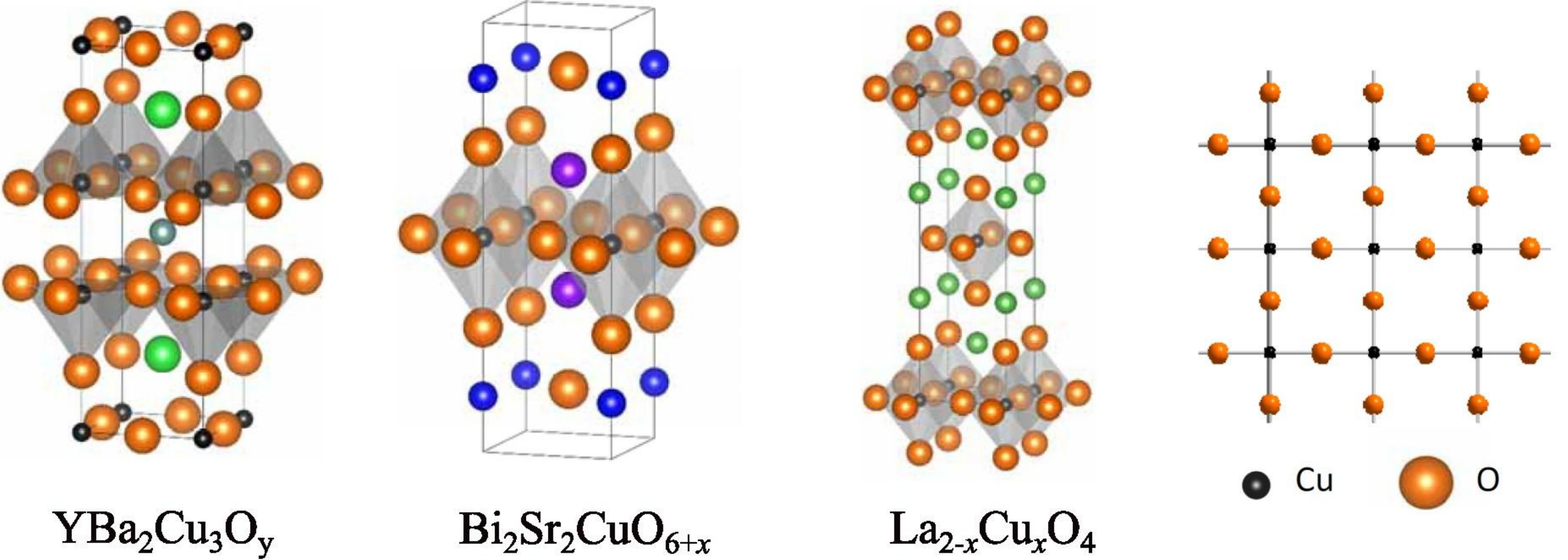}
	\caption{Schematic crystal structure of several representative cuprates
	}\label{}
\end{figure}
\begin{figure*}\center
	\includegraphics[width=0.8\linewidth]{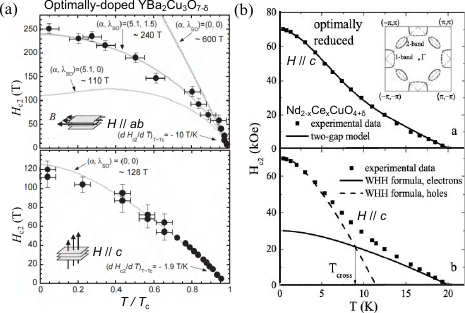}
	\caption{(a) \( H_{\text{c2}} (T) \) for optimally doped YBCO single crystal under \( H \parallel ab \) (parallel to the CuO$ _{2} $ plane, upper) and \( H \parallel c \) (perpendicular to the CuO$ _{2} $ plane, lower). The dashed lines correspond to the single-band WHH fitting curves \cite{sekitani2004upper}. (b) Temperature dependence of \( H_{\text{c2}} \) for an optimally reduced Nd\(_{1.85}\)Ce\(_{0.15}\)CuO\(_4\) film with \( H \parallel c \). The solid line represents the two-band model fitting. In the lower panel, the solid and dashed lines represent fits based on the electron single-band and hole single-band WHH models, respectively \cite{charikova2013upper}.
	}\label{}
\end{figure*}
\section{The upper critical field of Cuprate}

The discovery of cuprate high-temperature superconductors dates back to 1986, with Bednorz and Müller reporting superconductivity at a $T_{\rm{c}}$ exceeding 30 K in La$_{2-x}$Ba$_{x}$CuO$_{4}$. This groundbreaking discovery rapidly stimulated a worldwide effort to explore high-temperature superconductivity. One year later, superconductivity with $T_{\rm{c}}$ reaching 90 K was discovered in the Y-Ba-Cu-O system \cite{sekitani2004upper}, marking a significant milestone as the critical temperature exceeded the boiling point of liquid nitrogen (77 K). Cuprate high-temperature superconductors feature a layered perovskite structure with pronounced two-dimensional characteristics, consisting mainly of CuO$_2$ planes acting as conducting layers, interleaved with insulating layers serving as charge reservoirs, as illustrated in Fig. 2.
\subsection{The temperature dependence of $  H_{\rm{c2}} $ in cuprate}
For cuprate superconductors, the temperature dependence of $  H_{\rm{c2}} $ shows that the $ H_{\rm{c2}}^{ab}(T) $ curve displays a convex trend, whereas the $ H_{\rm{c2}}^{c}(T) $ curve exhibits either a more linear or concave dependence near $T_{\rm{c}}$, such as Sr$ _{1-x} $La$_{ x} $CuO$ _{2} $ \cite{PhysRevB.80.024501}, Pr$ _{2-x} $Ce$ _{x} $CuO$ _{4-y} $ (PCCO) \cite{wu2014superconducting}, Nd$ _{2-x} $Ce$ _{x} $Cu$ _{4+\delta} $ (NCCO) \cite{charikova2013upper,gantmakher2000temperature}, and Bi$ _{2} $Sr$ _{2} $CuO$ _{6+\delta} $ (2201) \cite{PhysRevB.73.014528}. Here, \(H_{\rm{c2}}^{ab}\) and \(H_{\rm{c2}}^c\) represent the upper critical fields when the magnetic field is applied in the \(ab\)-plane and along the \(c\)-axis, respectively.

Fig. 3(a) presents the temperature dependence of $ H_{\rm{c2}} $ for an optimally doped YBa$_{2}$Cu$_{3}$O$_{7-\delta}$ (YBCO) thin sample, measured using a contactless radio frequency transmission technique (RF method) \cite{sekitani2004upper}. The single-band WHH fitting can effectively describe the behavior of $ H_{\rm{c2}} (T) $. For $ H_{\rm{c2}}^{ab}(T) $, the Maki parameter $ \alpha_{\rm{M}} $ and spin-orbit scattering $ \lambda _{\rm{so}} $ obtained from the fitting are 5.1 and 1.5, indicating that the spin paramagnetic pair breaking effect is dominant. When ($ \alpha_{\rm{M}} $, $ \lambda _{\rm{so}} $) = (0, 0), the WHH model can also effectively fit the behavior of \( H_{\rm{c2}}^{c}(T) \), suggesting that the pair-breaking is solely caused by the orbital-limiting effect. In general, most cuprate superconductors are single-band systems that can be effectively characterized by the WHH model. However, in electron-doped superconducting NCCO single crystals \cite{gantmakher2000temperature} and thin films \cite{PhysRevB.58.11734,charikova2008effect,charikova2013upper}, the anomalous upward curvature observed in $ H_{\rm{c2}}^{c}(T) $ cannot be well described by the WHH model. When attempting to describe the $ H_{\rm{c2}}(T) $ using two independent WHH equations based on Eq. (11) with \( \lambda_{12} = 0 \), as shown in Fig. 3(b), it can be observed that the behavior results from a superposition of WHH curves for two types of carriers (electrons and holes), each associated with different critical fields, $ H_{\rm{c2}}^{1}$ and $ H_{\rm{c2}}^{2}$, as well as different transition temperatures, $T_{\rm{c}}^{1}$ and $T_{\rm{c}}^{2}$. Therefore, the two-band model effectively captures the essence of the $H_{\rm{c2}}(T)$ behavior for NCCO.

\begin{figure*}\center
	\includegraphics[width=0.8\linewidth]{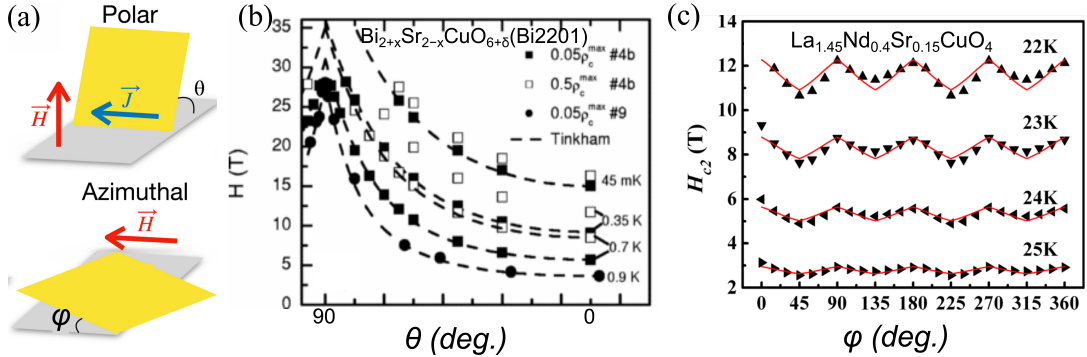}
	\caption{(a) Schematic of the measurement geometry relative to the sample (yellow square). (b) $ H_{\rm{c2}}(\theta) $ at different temperatures ranging from 45 mK to 0.9 K for Bi2201 \cite{PhysRevB.75.064512}. (c) $ H_{\rm{c2}}(\varphi) $ at different temperatures for La$ _{1.45} $Nd$ _{0.4} $Sr$ _{0.15} $CuO$ _{4} $ \cite{teng2013stripe}.
	}\label{}
\end{figure*}

\subsection{Anisotropy of $ H_{\rm{c2}}$ in cuprate}

Due to their unique layered structure, cuprate superconductors exhibit a considerable distance between the CuO$ _{2} $ planes and the charge reservoir layers, resulting in significant anisotropy. Fig. 4(a) presents a schematic illustration of the angles, where $ \varphi $ represents the rotation angle of the magnetic field within the \(ab\) plane, and \(\theta\) represents the angle between the magnetic field and the \(c\) axis. In Bi2201 superconductors, it can be observed that $  H_{\rm{c2}}$ exhibits a cusplike behavior at $ \theta=90$$^{\circ} $ (see Fig. 4(b)). Therefore, the 2D Tinkham model can be used to describe the angular dependence of $  H_{\rm{c2}}$. The dashed lines in Fig. 4(c) represent the Tinkham fit, which aligns well with the expected characteristics of conventional two-dimensional layered superconductors. In contrast to conventional superconductors, cuprates exhibit an anisotropic superconducting gap with $ d $-wave pairing symmetry. This is also reflected in their transport properties. The in-plane angular dependence of the upper critical field manifests as a fourfold symmetry, as observed in La$ _{1.45} $Nd$ _{0.4} $Sr$ _{0.15} $CuO$ _{4} $ (see Fig. 4(c)). 

\subsection{The doping dependence of $ H_{\rm{c2}} $ in cuprate}

\begin{figure*}\center
	\includegraphics[width=0.9\linewidth]{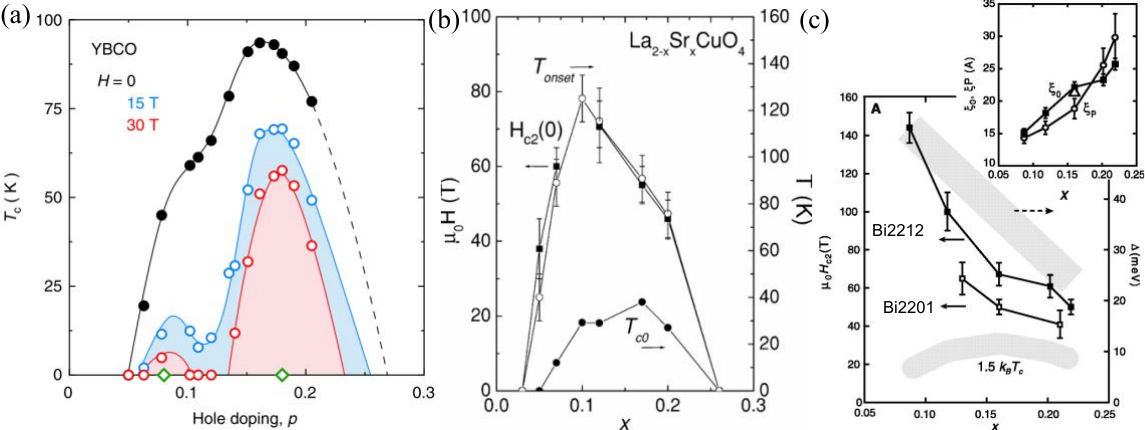}
	\caption{(a) $ T_{\rm{c}} $ of YBCO as a function of doping ($ p $) at different magnetic field values ($ H $) \cite{grissonnanche2014direct}.  $ T_{\rm{c}} $ is determined at zero resistance. Two peaks are observed in $  H_{\rm{c2}} $ and in superconducting phase diagram, located at $ p $=0.08 and $ p $=0.18. (b) \( H_{\rm{c2}}(0) \) as a function of \( x \) in LSCO (solid squares). \( H_{\rm{c2}}(0) \) values are extrapolated from measurements up to 45 T \cite{PhysRevB.73.024510}. (c) The $ x $ dependence of $  H_{\rm{c2}} $ in Bi 2212 (solid squares) and in Bi 2201 (open squares) measured by scaling of Nernst profiles.  The upper gray stripe represents the \( x \) dependence of the ARPES gap \(\Delta_0\), while the lower gray stripe represents $ T_{\rm{c0}} $. The inset shows a comparison of the coherence length \(\xi_0\) (solid squares) obtained from \( H_{\rm{c2}} \) and the Pippard length \(\xi_P\) (open circles) obtained from \(\Delta_0\) \cite{doi:10.1126/science.1078422}.
	}\label{}
\end{figure*}
Fig. 5 illustrates the \( H_{\mathrm{c}} \) of YBCO, LSCO, Bi2201, and Bi$ _{2} $Sr$ _{2} $CaCu$ _{2} $O$ _{8} $ (Bi2212) as a function of doping (\( p \)/$ x $) under different magnetic fields \cite{grissonnanche2014direct,PhysRevB.73.024510,doi:10.1126/science.1078422}. Clearly, the $  H_{\rm{c2}}(p) $ of YBCO and LSCO show variations that align with changes in $ T_{\rm{c}}(p) $. Additionally, the \( H \)–\( p \) phase diagram of superconductivity in YBCO features two distinct peaks, positioned at \( p_1 \approx 0.08 \) and \( p_2 \approx 0.18 \). Strong evidence indicates the presence of two distinct phase transitions in YBCO — one occurring at $ p_{1} $, and the other at a critical doping level corresponding to $ p_{2} $ \cite{tallon2001doping}. Notably, the Fermi surface is observed to undergo a transformation at $ p=0.08 $ and another near $ p\approx0.18 $ \cite{PhysRevB.83.054506}. Constructing the \( H \)–\( p \) (or \( H \)–\( x \)) phase diagram provides a way to identify the locations of these quantum critical points (QCPs). A similar behavior has also been observed in iron-based superconductors.
In contrast, for Bi2201 and Bi2212, different behaviors are observed in the $ H_{\rm{c2}} $ and $ T_{\rm{c}} $ as functions of doping concentration \cite{doi:10.1126/science.1078422}. The observed increase in $  H_{\rm{c2}} $ with decreasing doping concentration ($ x $) suggests that the superconducting pairing strength is enhanced in underdoped samples. This behavior contrasts with the trend of the superfluid density, which generally decreases as doping is reduced \cite{bonn1996surface,PhysRevLett.62.2317}. The increase in $ H_{\rm{c2}} $ indicates that the pairing potential remains robust in the underdoped regime, despite the reduced carrier density. It is believed that the superconducting dome is determined by both pairing strength and superfluid density.

\section{The upper critical field of FeSCs}
\begin{figure}\center
	\includegraphics[width=1\linewidth]{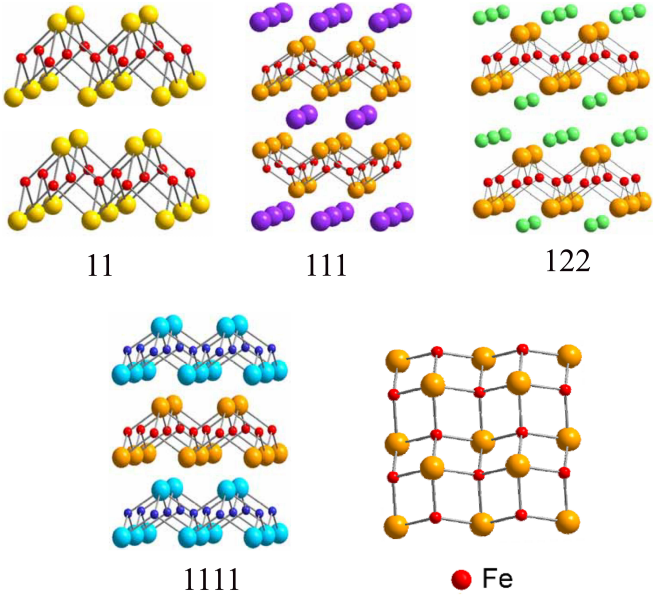}
	\caption{Schematic crystal structure of several representative FeSCs}\label{}
\end{figure}
\begin{figure*}\center
	\includegraphics[width=1\linewidth]{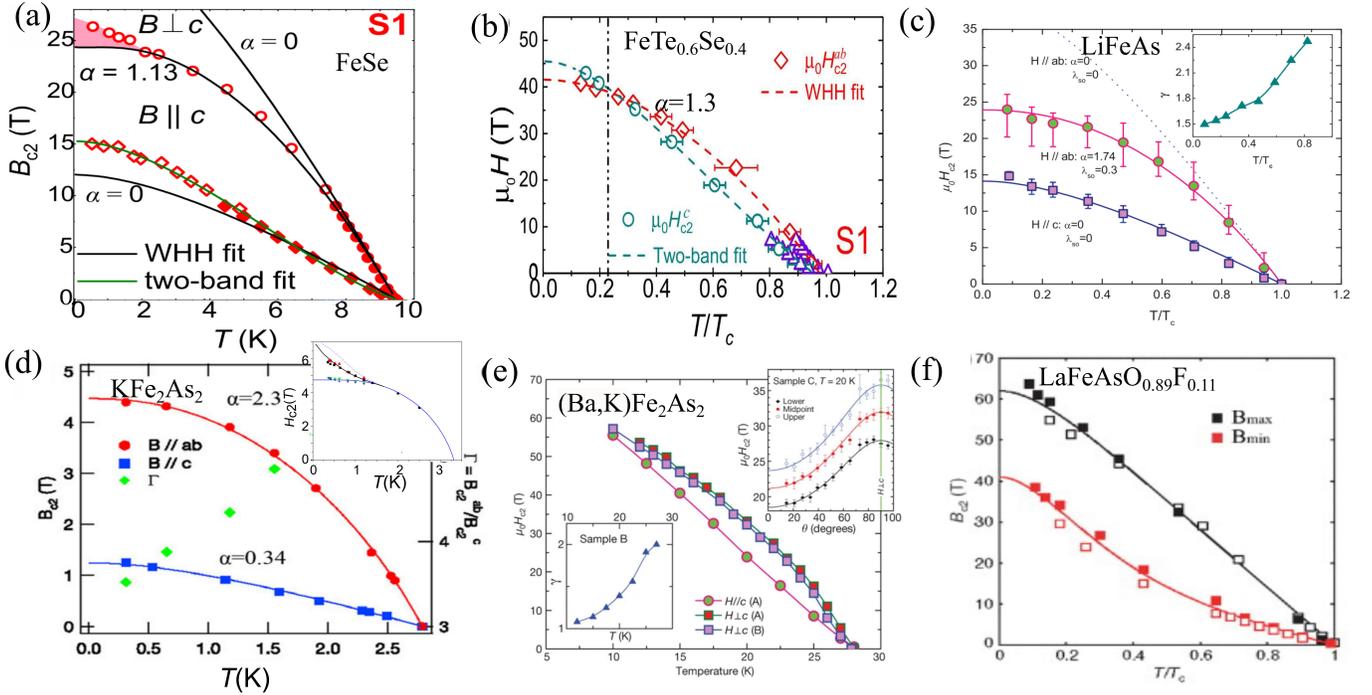}
	\caption{The temperature dependence of $H_{\rm{c2}}$ for FeSe \cite{PhysRevB.104.L140504}, FeTe$ _{0.6} $Se$ _{0.4} $ \cite{pan2024novel}, LiFeAs \cite{PhysRevB.83.174506}, KFe$ _{2} $As$ _{2} $ \cite{PhysRevLett.119.217002,terashima2009resistivity}, (Ba,K)Fe$ _{2} $As$ _{2} $ \cite{yuan2009nearly}, and LaFeAsO$ _{0.89} $F$ _{0.11} $ \cite{RN1488}.  The solid and dashed lines in the Figs. (a-c) and (e-f) represent fitted curves, either WHH fitting or two-band fitting.}\label{}
\end{figure*}

In 2008, the discovery of superconductivity in LaO$_{1-x}$F$_x$FeAs at 26 K stimulated the widespread interest in high-temperature FeSCs \cite{kamihara2008iron}. In comparison to cuprate superconductors, the layered crystal structure, the antiferromagnetic order in the parent compounds (though with different AFM configurations), and the distinct band structures make FeSCs an important platform for investigating high-temperature superconductivity mechanisms. So far, four main categories of FeSCs have been classified based on the crystal structures of their stoichiometric parent compounds, as illustrated in Fig. 6: the 11-type (FeTe, FeSe, FeS), the 111-type ($ A $FeAs, where $ A $ represents an alkali metal), the 122-type ($ Ae $Fe$_2$As$_2$, with $ Ae $ being Ba, Sr, K, or Ca), and the 1111-type ($ Ln $OFeAs, where $ Ln $ denotes a rare-earth element). High-temperature superconductivity originates from the FeSe/FeAs layers, similar to the CuO planes in cuprate high-temperature superconductors.

\subsection{The temperature dependence of $  H_{\rm{c2}} $ in FeSCs}
Iron-based superconductors exhibit relatively high $H_{\rm{c2}}$, with distinct temperature dependencies due to different contributions from multiband effects and spin-paramagnetic effects. Fig. 7 shows several typical temperature dependencies of $H_{\rm{c2}}(T)$ in iron-based superconductors. Specifically, in iron chalcogenide (FeCh) superconductors (FeSe) \cite{PhysRevB.104.L140504}, for $H \parallel c$, the temperature dependence of $H_{\rm{c2}}$($ T $) is nearly linear, with a slight upturn at low temperatures. The value of $H_{\rm{c2}}$ significantly exceeds the predicted upper limit based on the WHH theory \cite{werthamer1966temperature}, where both the Maki parameter $\alpha_{\rm{M}}$ and the spin-orbit interaction $\lambda$ are zero, as shown by the black lines in Fig. 7(a). In contrast, for $H \parallel ab$, the $H_{\rm{c2}}$($ T $) curve shows a convex shape, tending to saturate at low temperatures, with values falling below the WHH prediction at low temperatures when $\alpha$ = 0 and $\lambda$ = 0, indicating that the spin-paramagnetic effect cannot be neglected. Similarly, the $H_{\rm{c2}}(T)$ curves in Ba-122 \cite{yuan2009nearly} and LiFeAs \cite{PhysRevB.83.174506} exhibit comparable characteristics, with WHH fitting revealing relatively large Maki parameters, as illustrated in Figs. 7(c) and (e). However, in the FePn-1111 system (LnFeAs(O,F)) \cite{RN1488}, the behavior is notably different, as illustrated in Fig. 7(f). $ B_{\rm{min}}(T) $ and $ B_{\rm{max}}(T) $ represent $H_{\rm{c2}}^{c}(T)$ ($ H$ $\Arrowvert$ $c $) and $H_{\rm{c2}}^{ab}$ ($ H$ $\Arrowvert$ $ab $), respectively. $H_{\rm{c2}}^{c}(T)$ exhibits a linear increase as the temperature decreases near $ T_{\rm{c}} $, followed by a distinct upward curvature. This behavior can be explained by a two-band theory and is reminiscent of MgB$ _{2} $, which provided the first experimental evidence for two-band superconductivity in FeSCs \cite{hunte2008two}. The hypothesis of multi-band superconductivity was later confirmed by ARPES experiments \cite{ding2008observation}.



In FeSCs, WHH fitting typically yields a large Maki parameter, suggesting that the spin-paramagnetic effect is dominant. When $ \alpha _{\rm{M}} $ is sufficiently large and the superconductors are in the clean limit, the Fulde–Ferrell–Larkin–Ovchinnikov (FFLO) state may appear, leading to non-zero momentum of Cooper pairs and spatial oscillations of the superconducting order parameter. Fig. 7(a) clearly shows that in FeCh-FeSe superconductors, the Pauli limit is significantly exceeded, accompanied by an upward curvature in $H_{\rm{c2}}(T)$. Generally, for a conventional $ s $-wave superconductor, the FFLO state is known to be highly sensitive to nonmagnetic disorder, existing only in crystals under the clean limit conditions \cite{matsuda2007fulde,takada1970superconductivity}. For $H_{\rm{c2}}^{ab}$ in FeSe, an unusual superconducting phase was observed at low temperatures across all crystals, regardless of varying levels of disorder \cite{PhysRevB.104.L140504}. This finding indicates that the high-field superconducting phase in FeSe does not exhibit the typical characteristics of a conventional FFLO state. However, theoretical calculations suggest that the FFLO state may be less vulnerable to disorder in unconventional superconductors, such as disordered $ s $-wave \cite{PhysRevB.78.054501} or $ d $-wave superconductors \cite{PhysRevB.72.184501,ptok2010fulde}. In addition to the FFLO state, spin density wave (SDW) order has also been considered as a potential mechanism for the high-field superconducting phase in the heavy-fermion superconductor CeCoIn$ _{5} $. Theoretical calculations suggest that the SDW order near $  H_{\rm{c2}} $ is driven by a strong spin-paramagnetic pair-breaking effect combined with the nodal gap structure. In FeSe, both the electron-type band, which exhibits a small gap, and the hole-type band, with a larger gap, show gap nodes or deep minima, facilitating the possibility of field-induced SDW order. In addition, the multi-band effect in iron-based superconductors is also one of the possible origins of the upturn in the upper critical field. To better understand the origin of the high-field superconducting phase in FeSe, further studies, including nuclear magnetic resonance and neutron diffraction, are crucial.

In FePn-122 superconductors, an upward curvature in the upper critical field has also been observed, as shown in the inset of Fig. 7(d) \cite{PhysRevLett.119.217002}. Thermodynamic evidence for an FFLO state in KFe$ _{2} $As$ _{2} $ was obtained by investigating its $ H-T $ phase diagram near the Pauli limit using magnetic torque ($ \tau $) and specific heat ($ C_{p} $) measurements. 
Notably, only under strict alignment conditions does the magnetic field lead to a distinctive upturn in $H_{\rm{c2}}(T)$, surpassing the conventional Pauli limit. It also revealed a double transition, indicating a sequence from a homogeneous superconducting state to the FFLO state and finally to the normal state. This finding is crucial as it suggests the existence of a highly unconventional superconducting phase. 

\begin{figure*}\center
	\includegraphics[width=1\linewidth]{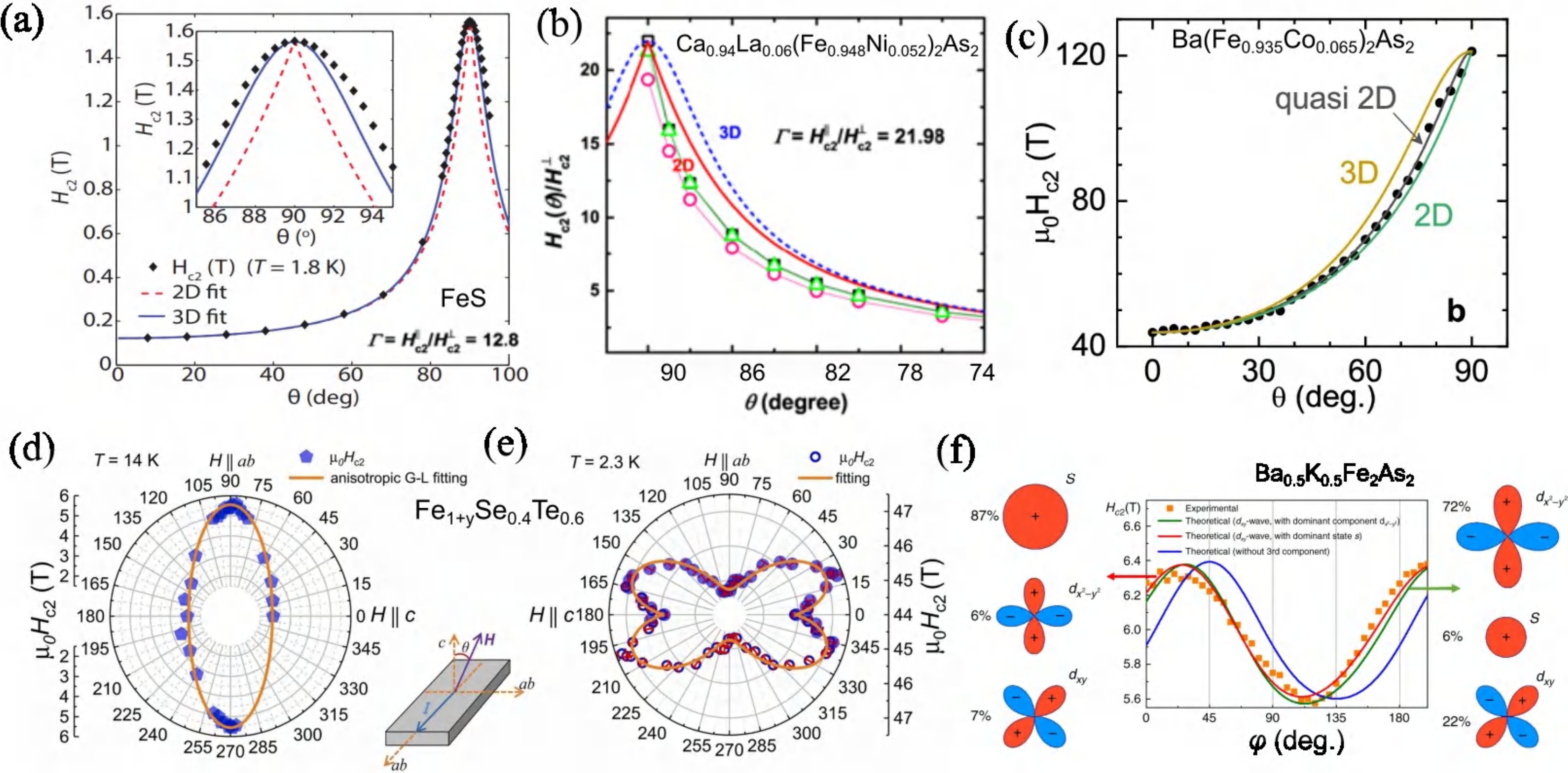}
	\caption{The angle dependence of the upper critical field $ H_{\rm{c2}}(\theta) $ for FeS \cite{borg2016strong}, Ca$ _{0.94} $La$ _{0.06} $(Fe$ _{0.948} $Co$ _{0.052} $)$ _{2} $As$ _{2} $ \cite{zhou2017anomalous}, and Ba(Fe$ _{0.935} $Co$ _{0.065} $)$ _{2} $As$ _{2} $ \cite{llovo2021multiband}. Angle dependent of $ H_{\rm{c2}} $ at $ T $= 14 K (d) and 2.3 K (e) for Fe$ _{1+y} $Se$ _{0.4} $Te$ _{0.6} $ \cite{pan2024novel}. (f) In-plane angle dependence of $ H_{\rm{c2}} (\varphi) $ at 38.4 K for Ba$ _{0.5} $K$ _{0.5} $Fe$ _{2} $As$ _{2} $ \cite{li2017nematic}.}\label{}
\end{figure*}

\subsection{Anisotropy of $H_{\rm{c2}}$ in FeSCs}

In FeCh-FeS superconductors, the angle dependence of $H_{\rm{c2}}$ at 1.9 K with respect to $ \theta $ is illustrated in Fig. 8(a) \cite{borg2016strong}. The 3D GL model provides a significantly better fit for the $H_{\rm{c2}}(\theta)$ curve compared to the 2D Tinkham model. The anisotropy parameter $\varGamma$ is 12.8, indicating a strongly anisotropic three-dimensional superconducting environment in FeS. However, in FePn-122 superconductors, a cusplike behavior near $ \theta $ = $0^{\circ} $ is observed in the $H_{\rm{c2}}(\theta)$ curves, as depicted in Figs. 8(b)-(c) \cite{zhou2017anomalous}. For FeCh-FeS, the interlayer distance is $\sim$ 5.034 $\rm\AA$, and the coherence length along the $ c $ axis ($\xi_c$) is around 343 $\rm\AA$. In contrast, FePn-Ca$ _{0.94} $La$ _{0.06} $(Fe$ _{0.948} $Ni$ _{0.052} $)$ _{2} $As$ _{2} $ has a $ c $-axis coherence length ($\xi_c$) ranging from about 3 to 16 $\rm\AA$ and for Ba(Fe$_{{0.935}}$Co$_{{0.065}}$)$_2$As$_2$, $\xi_c$ (0 K) is $\sim$ 10 $\rm\AA$. In Bi${_{2.2}}$Sr${_{2}}$Ca${_{0.8}}$Cu${_{2}}$O$_{8+\delta}$, the distance between adjacent superconducting layers reaches $ \sim $ 12 $\rm\AA$, significantly exceeding the coherence length $\xi_c$ ($ \sim $ 1.6 $\rm\AA$). These observations indicate that a large interlayer distance and a short coherence length along the $ c $-axis are crucial factors contributing to the occurrence of two-dimensional superconductivity \cite{zhou2017anomalous}. Furthermore, in the insets of Figs. 7(c) and 7(e), the temperature dependence of $ H_{\rm{c2}} $ anisotropy parameter for FePn-111 (LiFeAs) and FePn-122 ((Ba,K)Fe$ _{2} $As$ _{2} $) is also shown. The anisotropy ratio $\varGamma$ trends towards isotropy as the temperature decreases. (Ba,K)Fe${_{2}}$As${_{2}}$ was the first iron-based superconductor found to be near isotropy. This discovery is particularly significant, as it demonstrates that its superconducting properties increasingly resemble 3D behavior at lower temperatures, which is beneficial for potential applications.

In Fig. 7(b), a crossover between the $H_{\rm{c2}}^{c}(T)$ and $H_{\rm{c2}}^{ab}(T)$ curves for FeSe$ _{0.6} $Te$ _{0.4} $ is observed \cite{pan2024novel}. This crossover is notably robust against disorder. Similar unusual behavior has also been reported in other compounds, including Ba$_{1-x}$K$_x$Fe$_2$As$_2$ \cite{yuan2009nearly}, $ A $Cr$_3$As$_3$ \cite{PhysRevB.100.214512}, and $ A _2$Cr$_3$As$_3$ \cite{PhysRevB.91.220505,cao2018superconductivity}, where $ A $ represents an alkali metal. Figs. 8(d)-(e) present the angular dependence of $ H_{\rm{c2}} $ in FeSe$ _{0.6} $Te$ _{0.4} $ at 14 K (above the crossover point) and 2.3 K (below the crossover point) for the out-of-plane direction, respectively \cite{pan2024novel}. The behavior of $H_{\rm{c2}}(\theta)$ above and below the crossover point in FeSe$_{0.4}$Te$_{0.6}$ exhibits an evolution from twofold (C2) to fourfold (C4) symmetry. 
A spin-orientation-locking model in quasitwo-dimensional superconductors is proposed to explain the observed behavior. This model introduces the concept of half-itinerant carriers, whose spins are confined within the $ ab $-plane, while both charge and spin remain itinerant. For $  H$ $\Arrowvert$ $ab $-plane, an angle between the magnetic field and the spins in a Cooper pair leads to additional magnetization energies of $ +\mid M \cdot H \mid $ and $ -\mid M \cdot H \mid $, leading to a peak in depairing energy. When the magnetic field is applied along the $ c $-axis, perpendicular to the spin directions, the magnetization energy remains zero,  and no spin-paramagnetic depairing effects occur. Furthermore, both orbital depairing and spin-paramagnetic depairing effects depend on the angle $ \theta $. As the magnetic field rotates from $H$ $\parallel$ $c$ to $H$  $\parallel$ $ab$, the rate of attenuation for orbital depairing effects surpasses the rate of enhancement for in-plane spin-paramagnetic depairing effects. However, the general applicability of this theoretical model requires further validation across different types of iron-based superconductors.

For FePn-122, Fig. 8(c) clearly shows that the out-of-plane angular dependence of $ H_{\rm{c2}} $ exhibits C2 symmetry, consistent with the typical characteristics of layered superconductors. Fig. 8(f) illustrates the in-plane angular dependence of $H_{\rm{c2}}$ at 38.4 K (close to $T_c \approx 39$ K) for FePn-122 (Ba$_{{0.5}}$K$_{{0.5}}$Fe$_2$As$_2$) \cite{li2017nematic}. In the superconducting phase diagram of Ba$_{{1-x}}$K$_{{x}}$Fe$_2$As$_2$, the nematic phase can be observed and persists until it vanishes at $x \approx 0.3$. Electronic nematic order is commonly observed in other unconventional superconductors, including high-$ T_{\rm{c}} $ cuprates and heavy-fermion compounds. It is characterized by the breaking of tetragonal rotational symmetry, often leading to a structural transition to an orthorhombic phase. Based on the behavior of $H_{\rm{c2}}(\phi)$, it is evident that the C2 symmetry of the superconducting condensate persists from the onset of superconductivity to low temperatures and across a range of dopings around $x = 0.5$, indicating that the nematic state is present over a broad doping range. Notably, this nematic order is not accompanied by the typical tetragonal symmetry breaking of the crystal lattice or the emergence of magnetic order, making it an unusual characteristic among iron-based superconductors. The unusual nematic state is thought to arise from the weak coupling between the quasi-degenerate $ s $±-wave and $ d $-wave components in the superconducting condensate.

\subsection{The doping dependence of $  H_{\rm{c2}} $ in FeSCs}
\begin{figure}\center
	\includegraphics[width=1\linewidth]{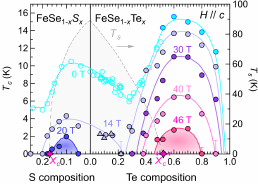}
	\caption{Superconducting phase diagram of FeSe$ _{1-x} $(Te/S)$ _{x} $ under different magnetic fields. The gray dashed line indicates zero-temperature intercepts of the structural transition $T_{\rm{s}}$, and $x_{\rm{c}}$ represents the nematic quantum critical points \cite{PhysRevX.13.011032}.}\label{}
\end{figure}
Fig. 9 presents the $T_{\rm{c}}$ phase diagram of FeSe$ _{1-x} $(Te/S)$ _{x} $ under different magnetic fields \cite{PhysRevX.13.011032}. At $ H $ = 0, the doping of S and Te results in the appearance of two distinct superconducting domes. The nematic transition temperature $T_{\rm{s}}$ gradually decreases with increasing doping and eventually vanishes around $x(\rm{S}) \sim 0.17$ and $x(\rm{Te}) \sim 0.5$. Systematic elastoresistivity measurements on FeSe$_{1-x}$Te$_{x}$ single crystals identified a non-magnetic nematic QCP at $x \sim 0.5$, isolated from any other long-range orders, located at the center of the superconducting dome \cite{ishida2022pure}. As the magnetic field increases, the superconducting dome for FeSe$_{{1-x}}$Te$_x$ contracts into a narrower dome around the nematic QCP. Similar behavior has also been observed in cuprate superconductors, as mentioned in the previous section. Analysis of $H_{\rm{c2}}$ reveals that the Pauli-limiting field increases as the system nears the QCP, indicating that the pairing interaction is significantly enhanced by nematic fluctuations emanating from the QCP. The dependence of $ H_{\rm{c2}} $ on doping plays a crucial role in exploring QCPs. However, FeSe$ _{1-x} $S$ _{x} $ exhibits a QCP associated with magnetic ordering. Unlike FeSe$_{1-x}$Te$_x$, the superconducting dome in FeSe$ _{1-x} $S$ _{x} $ does not contract around the QCP as the magnetic field increases, likely due to the influence of spin fluctuations. Further research is essential to achieve a more profound understanding.

\section{The upper critical field in infinite-layer nickelate superconductors}
In 2021, superconductivity was first observed in infinite-layer nickelate films — Nd$_{1-x}$Sr$_{x}$NiO$_{2}$ \cite{RN1471}. Subsequently, the development of rare-earth ($ R $) variants, substituting Nd with Pr or La, has led to the emergence of a new family of superconducting infinite-layer nickelates. The films undergo a topotactic transition from the perovskite phase to the infinite-layer phase, as shown in Fig. 10. These infinite-layer nickelate superconductors are not only isostructural to the well-known cuprate superconductor CaCuO$_2$, but also feature Ni and Cu with a similar formal 3$ d $$^9$ electronic configuration in their respective parent compounds.
\subsection{The temperature dependence of $  H_{\rm{c2}} $ in infinite-layer nickelate superconductors}
\begin{figure}\center
	\includegraphics[width=1\linewidth]{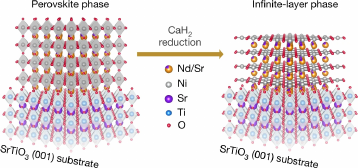}
	\caption{\textbf{(a)} Topotactic reduction of nickelate thin films \cite{RN1471}}\label{}
\end{figure}
\begin{figure*}\center
	\includegraphics[width=1\linewidth]{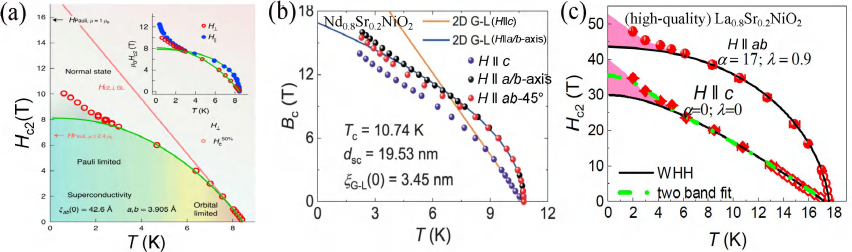}
	\caption{Temperature dependence of \(H_{c2}\) for \(H \parallel c\) and \(H \parallel ab\) in \(\text{\textit{R}}_{1-x}\text{Sr}_{x}\text{NiO}_2\)($ R $ = Nd(a-b), La(c)), determined using a 50\% $\rho_{\rm{n}} $ criterion \cite{RN1445,RN1466,PhysRevB.107.L220503}.
	}\label{}
\end{figure*}
The temperature dependence of  $H_{\rm{c2}}$ for $ R _{1-x} $Sr$ _{x} $NiO$ _{2} $ thin films was investigated by measuring resistivity in both inter-plane (\(H \parallel c\)) and in-plane (\(H \parallel ab\)) directions under magnetic fields \cite{RN1445,PhysRevB.107.L220503,wang2023effects}. As shown in Fig. 11, the $H_{\rm{c2}}(T)$ behavior of nickelate superconductors containing different rare-earth elements exhibits several common characteristics: (1) anomalous low-temperature upturn; (2) square-root temperature dependence of $H_{\rm{c2}}^{ab}$; (3) linear temperature dependence of $H_{\rm{c2}}^{c}$. Additionally, notable differences in anisotropy have also been observed.

The linear temperature dependence of $H_{\rm{c2}}^{c}$ in all nickelate thin films can be interpreted as indicating a dominant orbital depairing effect near $ T_{\rm{c0}} $, which fits well with the two-band model. Similar behavior has also been observed in most iron-based superconductors \cite{PhysRevB.83.174506,PhysRevB.81.020509,PhysRevB.81.184511,yuan2009nearly}. For \(H_{\rm{c2}}^{ab}\), a \((T_{\rm{c}} - T)^{1/2}\) temperature dependence is observed in La-, Pr-, and Nd-nickelate superconductors, which is similar to the behavior predicted by the GL model for thin films \cite{harper1968mixed}:
\begin{eqnarray*}
	\\H_{\rm{c2}}^{ab}(T)=\frac{\sqrt{12}\phi_{0}}{2\pi\xi_{ab}(0)d_{f}}(1-\frac{T}{T_{\rm{c0}}})^{1/2},
\end{eqnarray*}
where $ d_{f} $ is the film thickness. However, the deduced superfluid thickness shows an unphysical discrepancy consistently across all samples, indicating that the orbital depairing effect alone cannot fully explain the temperature dependence of \(H_{\rm{c2}}^{ab}\). This necessitates consideration of the paramagnetic depairing effect. Within the pair-breaking framework proposed by Abrikosov-Gorkov \cite{abrikosov1960contribution,tinkham2004introduction}, $ \ln({T}/{T_{\rm{c0}}} ) = \psi( {1}/{2} ) - \psi( {1}/{2} + {\alpha _{\rm{M}}}/{2\pi k_B T_{\rm{c0}}} ) $. 
When considering only the orbital depairing effect, it is characterized by $ \alpha _{\rm{M}}=D^{ab}eH/c $, where \(e\) is the electronic charge, \(c\) is the speed of light, and \(D^{ab}\) is the in-plane electronic diffusion coefficient. The \(H_{\rm{c2}}^{c}\)($ T $) derived from this term indeed exhibits a linear temperature dependence near \(T_{c0}\). On the other hand, when considering only the paramagnetic effect, it can be described by $ \alpha _{\rm{M}}=\frac{\tau_{\rm{so}}e^{2}\hbar H^{2}}{2m_{eff}^{2}} $\cite{tinkham2004introduction}, where $ \tau_{\rm{so}} $ is the spin-orbit scattering time and $ m_{\rm{eff}} $ is the electronic effective mass. Maki and Tsuneto \cite{maloney1972superconducting} derived an approximation for the temperature dependence of \(H_{\rm{c2}}\) in the limit of $ T-T_{\rm{c0}} \ll T_{\rm{c0}}$, expressed as \(-\ln(T/T_{\rm{c0}}) = 7 \zeta(3) ( \mu_{\rm{eff}} H / 2 \pi T_{c0})^2\), $ (\mu_{\rm{eff}} H)/(2 \pi T_{\rm{c0}})\ll 1 $, which exhibits a \((T_{\rm{c}} - T)^{1/2}\) dependence near \(T_{\rm{c0}}\). Here, \(\zeta\) is the Riemann zeta function, and \(\mu_{\rm{eff}}\) represents the effective magnetic moment. This further validates our analysis of the temperature dependence behavior of \(H_{\rm{c2}}\) in nickelate superconductors.

To reveal the intriguing behavior of the low-temperature upturn, it is necessary to verify it in nickelate thin films under the clean limit. La$_{0.8}$Sr$_{0.2}$NiO$_{2}$ thin films, prepared by the MBE method and exhibiting a macroscopic clean limit, possess a high superconducting transition temperature ($T_{\rm{c}}^{\rm{onset}} = 18.8$ K) and very large upper critical fields ($H_{\rm{c2}}^{c} (0)\approx 40$ T, $H_{\rm{c2}}^{ab}(0) \approx 52$ T), making them well-suited for studying this issue. 
By fitting using the WHH model, it is found that the similar low-temperature upturn behavior and a large parameter $ \alpha^{\rm{Maki}} $ were found in all infinite-layer nickelate superconductors. Additionally, the scaling curve of $H_{\rm{c2}}/H_{\rm{c2}}^{\text{WHH}} (0$ K$)$ in La$_{0.8}$Sr$_{0.2}$NiO$_{2}$ films with different $T_{\rm{c}}$ (i.e., different levels of disorder) shows nearly identical upturn behavior (see Fig. 12(d)), indicating that the anomalous upturn is independent of film quality and rare-earth elements.

\begin{figure*}\center
	\includegraphics[width=1\linewidth]{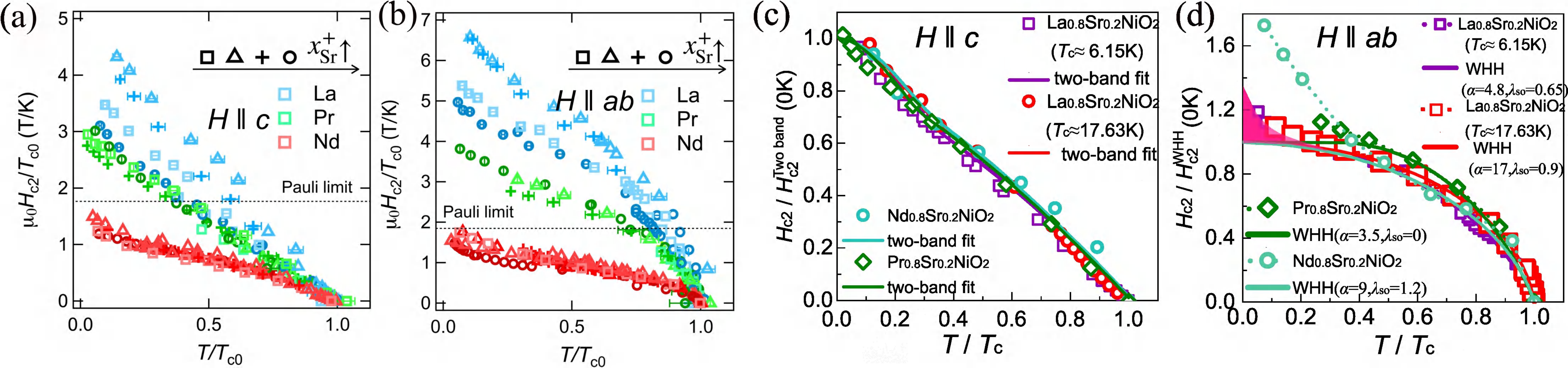}
	\caption{(a-b) Normalized $H_{\rm{c2}}$/$ T_{\rm{c0}} $ against reduced temperature $ T $/$ T_{\rm{c0}} $ of $R_{1-x}$Sr$_{x}$NiO$_{2}$ for $H \parallel c$ and $H \parallel ab$, respectively. Here, $ x $ = 0.15, 0.175, 0.2, and 0.225 \cite{PhysRevResearch.3.L042015,RN1466,wang2023effects} for Nd, $ x $ = 0.16, 0.18, 0.20, and 0.24 for Pr, and $ x $ = 0.15, 0.16, 0.18, and 0.20 for La. The dashed lines indicate the Pauli limit of $ H_{\rm{c2}} $ = 1.86 $ T_{\rm{c0}} $ \cite{wang2023effects}; (c-d) Normalized $H_{\rm{c2}}$/$H_{\rm{c2}}^{\rm{Two-band}}$(0 K) and Normalized $H_{\rm{c2}}$/$H_{\rm{c2}}^{\rm{WHH}}$(0 K) data for $R_{0.8}$Sr$_{0.2}$NiO$_{2}$ ($ R $ = La, Pr, or Nd) with different rare-earth elements \cite{PhysRevB.107.L220503}.
	}\label{}
\end{figure*}
The anomalous upturns observed in heavy-fermion superconductors \cite{matsuda2007fulde,kitagawa2018evidence}, organic superconductors \cite{singleton2000observation}, and some iron-based superconductors have been attributed to the FFLO state. However, for conventional $ s $-wave superconductors, the FFLO state is easily destroyed by impurities \cite{matsuda2007fulde}. Moreover, due to the fact that the film is in the dirty limit, the anomalous upturn behavior observed in nickelates is unlikely to be attributed to the FFLO state. What, then, is the cause of the low-temperature high-field phase? The high-field superconducting phase may originate from the presence of a spin-density wave (SDW). For instance, in the heavy-fermion superconductor CeCoIn$_{5}$, an additional SDW coexists with superconductivity \cite{kenzelmann2008coupled,kenzelmann2010evidence}. The hybridization of La 5$d$ and Ni 3$d$ orbitals enhances the interlayer electron coupling as the temperature decreases, which could induce SDW fluctuations. However, more evidence is needed to confirm the coexistence of SDW and superconductivity. The multiple-band effect could also be a reason for the low-temperature upturn in $ H_{\rm{c2}}(T)$.
For La-, Pr- and Nd-nickelates, the distinctions originate from the characteristics of 4$f$ electrons in the lattice: La$^{3+}$ has no 4$f$ electrons; Pr$^{3+}$, with a singlet ground state, is non-magnetic, similar to La$^{3+}$; while Nd$^{3+}$ is magentic. Fig. 12(a-b) presents the normalized $H_{\rm{c2}}$($T$) curves of these $R_{1-x}$Sr$_{x}$NiO$_{2}$ ($ R $ = La, Pr, Nd) nickelate superconductors \cite{wang2023effects}. 
Regardless of the specific doping level, $H_{\rm{c2}}$ of Nd-nickelate is significantly lower than that of La- and Pr-nickelates. Also noteworthy is that the $H_{\rm{c2}}$ of La- and Pr-nickelates greatly exceeds the conventional Pauli limit by approximately 3.5 times ($H_{\rm{c2}} \approx 1.86T_{\rm{c0}}$). Furthermore, the $H_{\rm{c2}}$ of Nd- and Pr-nickelates can be scaled onto a single curve through normalization. 
These observed differences are likely influenced by the presence of 4$ f $ electrons. The angular dependence of magnetoresistance measurements in La-, Pr-, and Nd-nickelates, as shown in Fig. 13, may provide further insights into the reasons behind the variations in $H_{\rm{c2}}$ among these materials. 

For $R_{1-x}$Sr$_{x}$NiO$_{2}$ ($ R $ = La, Pr, Nd), the magnetoresistance in both the normal state and fully superconducting state shows no dependence on the polar or azimuthal angles. Thus, only the superconducting transition region is considered. In this region, La- and Pr-nickelates exhibit a typical polar angular dependence. In Figs. 13(a)-13(b), the "$ \infty $"-shaped dependence on $ \theta $ suggests that superconductivity is less suppressed (lower resistance) when the external magnetic field is applied in the plane of NiO$ _{2} $, indicating a weaker depairing effect—this is a common characteristic of layered superconductors \cite{ihara1997make,iye1991anisotropy}. The "8"-shaped, two-fold (C2) symmetric $ \phi $ dependence shown in Figs. 13(e)-13(f) is also consistent with previous studies of other type-II superconductors \cite{iye1991anisotropy,iye1990dissipation}. In contrast, Nd-nickelate exhibits distinct behavior. The "8"-shaped $ \theta $ dependence in Fig. 13(c) indicates that superconductivity is more significantly suppressed by the in-plane magnetic field. Additionally, the four-fold (C4) symmetric $ \phi $ dependence in Fig. 13(g) suggests a potential influence of localized magnetic moments\cite{wang2023effects}. The anomalous $ \theta $ dependence in Nd-nickelate can be attributed to the combined effect of two mechanisms: the conventional anisotropic orbital depairing (AOD) effect due to the layered crystal structure, and the enhanced magnetic permeability (EMP) effect resulting from the anisotropy of the Nd$ ^{3+} $4$ f $ magnetic moment (Figs. 13(i)-13(j)), which is unique to Nd-nickelate) \cite{wang2023effects}. The AOD effect accounts for the $ \theta $ dependence of magnetoresistance observed in La- and Pr-nickelate, exhibiting a minimum resistance with in-plane magnetic fields. Regarding the EMP effect, the magnetic susceptibility of Nd$ ^{3+} $ exhibits a clover-leaf pattern, leading to maximum amplification of the local in-plane magnetic field, thereby enhancing the depairing effect and suppressing superconductivity. The investigation into the magnetoresistance of infinite-layer nickelate Nd${_{0.8}}$Sr${_{0.2}}$NiO$_2$ superconducting films \cite{RN1445} not only revealed an azimuthal clover-leaf pattern but also identified an additional two-fold (C2) symmetric component in the $ R(\phi) $ curves as the system approached low temperatures and high magnetic fields.

\begin{figure*}\center
	\includegraphics[width=0.8\linewidth]{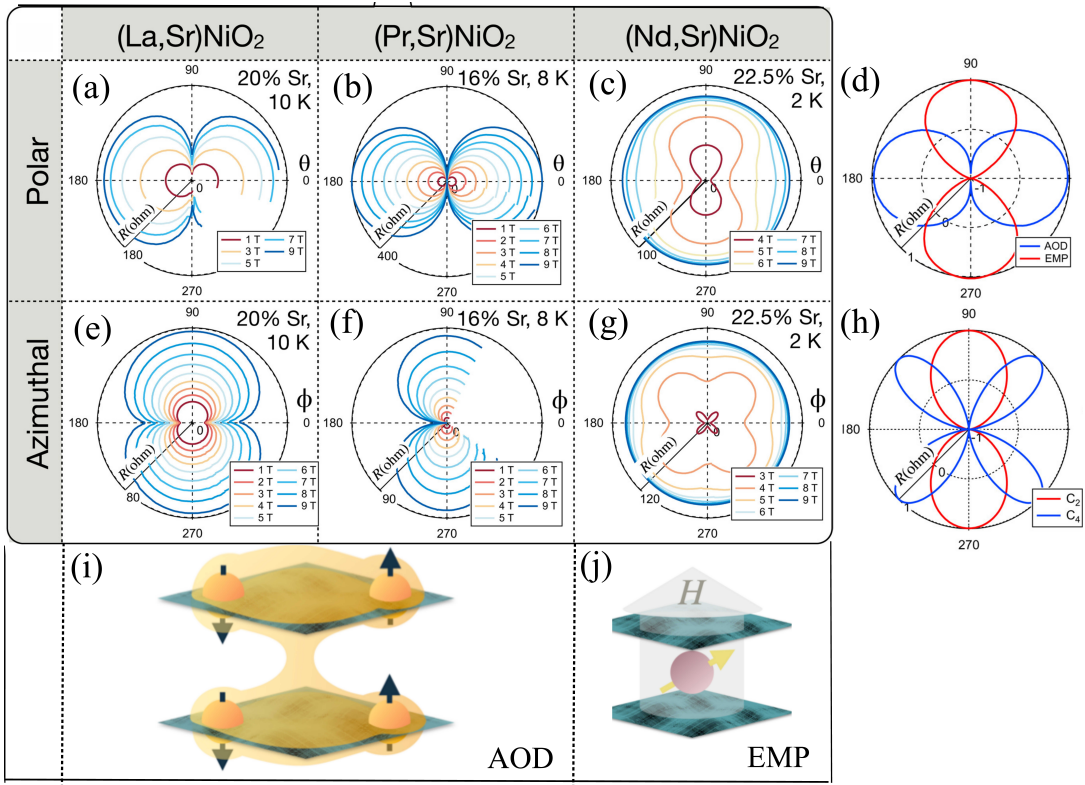}
	\caption{ Polar (a-c) and azimuthal (e-g) angle dependence of magnetoresistance in the superconducting transition region for La-, Pr-, and Nd-nickelates. (d and h) Fitting functions representing the AOD and EMP effects for $ \theta $- and $ \phi $-dependence. Schematic representations of (i) anisotropic orbital depairing (AOD) and (j) enhanced magnetic permeability (EMP) \cite{wang2023effects}.
	}\label{}
\end{figure*}

\subsection{Anisotropy of $H_{\rm{c2}}$ in infinite-layer nickelate superconductors}

Harold Y. Hwang discovered that $H_{\rm{c2}}$ is surprisingly isotropic at low temperatures in \(\text{Nd}_{0.775}\text{Sr}_{0.225}\text{NiO}_2\) thin films with $ T_{\rm{c}} $ of 8.5 K, despite the layered crystal structure, quasi-2D band structure ($d$$_{x^{2}-y^{2}}$) and the thin film geometry (as shown in Fig. 11(a)) \cite{RN1466}. Interestingly, the observed isotropy in \(\text{Nd}_{0.775}\text{Sr}_{0.225}\text{NiO}_2\) differs from the anisotropic $ H_{\rm{c2}}(T) $ behavior observed in infinite-layer nickelate superconductors, such as the high-quality \(\text{La}_{0.8}\text{Sr}_{0.2}\text{NiO}_2\) thin film. In Fig. 14(c), the \(H_{\rm{c2}}\) anisotropy (\(\gamma = H_{\rm{c2}}^{ab} / H_{\rm{c2}}^c\)) decreased monotonically from \(\sim 10\) near \(T_{\rm{c}}\) to \(\sim 1.5\) at 2 K in a high-quality \(\text{La}_{0.8}\text{Sr}_{0.2}\text{NiO}_2\) thin film with \(T_{\rm{c}}^{\rm{onset}} = 18.8 \, \text{K}\). The $ \gamma $ value in high-quality thin film is somewhat larger compared to previous reports on lower-quality thin films with lower $ T_{\rm{c}} $ (7 for La$ _{0.8} $Sr$ _{0.2} $NiO$ _{2} $ and 6 for La$ _{0.8} $Ca$ _{0.2} $NiO$ _{2} $ near $ T_{\rm{c}} $) \cite{wang2023effects}. For \(\text{Nd}_{0.775}\text{Sr}_{0.225}\text{NiO}_2\), the isotropy can be attributed to the easy-plane magnetic anisotropy of the Nd$ ^{3+} $ 4$ f $ moment, which increases the in-plane magnetic susceptibility, enhancing the suppression of superconductivity. First-principles calculations reveal that in LaNiO$ _{2} $, two small electron-like pockets around the $ \varGamma $ and $ A $ points are mainly derived from La orbitals, and with approximately 20\% hole doping, only a reduced 3D electron pocket remains at the $ A $ point. Therefore, the anisotropy seen in \(\text{La}_{0.8}\text{Sr}_{0.2}\text{NiO}_2\) thin films may be intrinsic to infinite-layer nickelates, unaffected by Nd$ ^{3+} $ magnetism, while orbital hybridization at lower temperatures increases electron coupling, thus reducing anisotropy.

\begin{figure*}\center
	\includegraphics[width=0.9\linewidth]{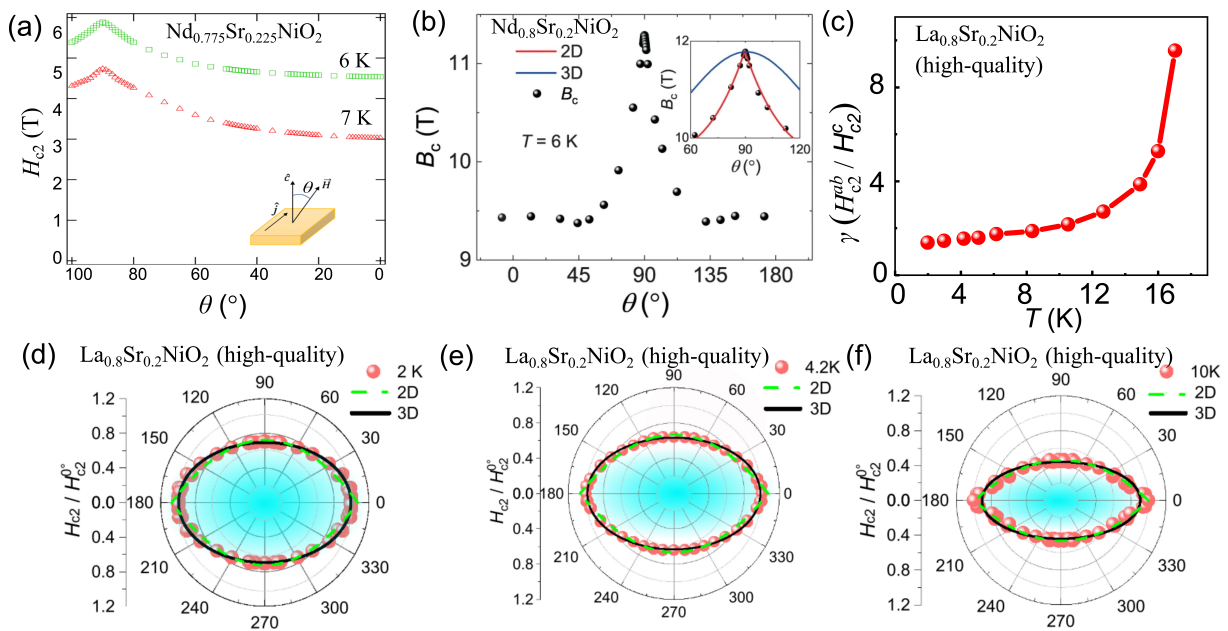}
	\caption{(a-b) Polar angle dependence of $ H_{\rm{c2}}$ in \(\text{Nd}_{0.775}\text{Sr}_{0.225}\text{NiO}_2\) and Nd$ _{0.8} $Sr$ _{0.2} $NiO$ _{2} $ \cite{RN1445,RN1466}. Inset of (a) shows a schematic definition of the magnetic field orientation angle $ \theta $. (c) Anisotropy of $ H_{\rm{c2}}$ (\(\gamma = H_{\rm{c2}}^{ab} / H_{\rm{c2}}^c\)) for the high-quality \(\text{La}_{0.8}\text{Sr}_{0.2}\text{NiO}_2\) thin film \cite{PhysRevB.107.L220503}. (d-e) Polar angle dependence of the normalized $ H_{\rm{c2}} (\theta)/H_{\rm{c2}}^{0^{\circ}} $, for the high-quality \(\text{La}_{0.8}\text{Sr}_{0.2}\text{NiO}_2\) thin film at various temperatures \cite{PhysRevB.107.L220503}. 
	}\label{}
\end{figure*}

Angular dependence of $ H_{\rm{c2}} $ can also provide valuable information about a superconductor’s depairing processes and dimensionality. As shown in Fig. 14(a), although angular modulation is clearly present in \(\text{Nd}_{0.775}\text{Sr}_{0.225}\text{NiO}_2\), the magnitude of $H_{\rm{c2}}$ remains predominantly isotropic and becomes more pronounced at lower temperatures \cite{RN1466}. Figs. 14(d-f) present the angle dependence of the normalized $H_{\rm{c2}}$ at 2 K, 4.2 K, and 10 K for a high-quality \(\text{La}_{0.8}\text{Sr}_{0.2}\text{NiO}_2\) thin film \cite{PhysRevB.107.L220503}. By fitting the data using both the 3D GL model and the 2D Tinkham model, a dimensional crossover from 2D to 3D superconductivity with decreasing temperature is observed. Such a crossover may be attributed to the hybridization of rare-earth 5$ d $ and Ni 3$ d $ orbitals, forming interstitial-$ s $ orbital, which enhances the coupling strength between the NiO\(_2\) planes and the rare-earth spacer layers at lower temperatures \cite{PhysRevB.102.220501,gu2020substantial}. However, all infinite-layer nickel-based 112 thin films studied thus far remain in the dirty limit, and further improvements in sample quality could help clarify the issue of $ H_{\rm{c2}} $ anisotropy.

\section{Conclusions and outlook}

The study of $H_{\rm{c2}}$ is crucial to the mechanism of the superconducting pair-breaking mechanism and the symmetry of superconducting pairing, which is mainly reflected in the following aspects: 

(i) The study of the temperature dependence of $  H_{\rm{c2}}(T) $ is crucial for understanding the fundamental physics of superconductivity, as it provides insights into the pairing mechanisms and limiting factors that govern superconducting behavior. Near $ T_{\rm{c}} $, the orbital depairing effect, which arises from the Lorentz force acting on Cooper pairs in the presence of a magnetic field, is the dominant factor limiting $  H_{\rm{c2}} $. In contrast, at low temperatures and under high magnetic fields, the spin-paramagnetic depairing effect becomes the primary constraint, where the Zeeman energy disrupts the Cooper pairs. Furthermore, unconventional superconducting states, such as the spin-triplet $ p $-wave pairing state \cite{PhysRevLett.108.057001}, the non-zero momentum FFLO state most in FeSCs \cite{PhysRev.135.A550,larkin1965nonuniform,PhysRevLett.124.107001}, the strongly correlated Bose-Einstein - Condensation (BEC) superconducting state \cite{kuchinskii2017temperature}, and multi-band superconducting states \cite{PhysRevB.67.184515}, can significantly affect the temperature dependence of $H_{\rm{c2}}$, even leading to the appearance of high-field phases in the low-temperature region \cite{kenzelmann2008coupled,PhysRevLett.109.027003,doi:10.1073/pnas.1413477111}.

(ii) Anisotropy of $H_{\rm{c2}}$. As a fundamental parameter, the superconducting anisotropy \(r = H_{\rm{c2}}^{ab} / H_{\rm{c2}}^c\) plays a key role in understanding the superconducting mechanism and its potential applications. 
The anisotropy of $H_{\rm{c2}}$ directly reflects the dimensionality of superconductivity. 
The in-plane anisotropy of $H_{\rm{c2}}$ is closely related to the structure of the superconducting gap and pairing symmetry. For instance, in cuprete superconductors, the in-plane $H_{\rm{c2}}$ shows fourfold symmetry ($ d $-wave) \cite{PhysRevB.54.R776}, 
and in nickelate superconductors, there is mixed symmetry (possibly $ s+d $-wave) \cite{RN1445}. Additionally, when the superconducting state is influenced by factors such as charge order, nematic phases, or strong spin-orbit coupling, as seen in FeSCs, changes in the symmetry of the superconducting wave function can also be reflected in the angular dependence of $H_{\rm{c2}}$. 

(iii) Doping dependence of $H_{\rm{c2}}$. The variation of $H_{\rm{c2}}$ with doping concentration can be used to track changes in the strength of the pairing potential, where higher $H_{\rm{c2}}$ corresponds to the stronger pairing potential and a smaller pairing size. On the other hand, the variation of $H_{\rm{c2}}$ with doping or pressure can directly reflect the relationship between superconductivity and other electronic ordered states, helping to precisely determine the location of the quantum critical point (QCP) and providing clues for understanding the superconducting pairing mechanism.

The analysis of  across these superconductors deepens our understanding of the superconducting pairing mechanisms and pairing symmetry, providing valuable directions for future research.
\color{black}

\section*{Acknowledgements}
\acknowledgements
This work was partly supported by 
the National Natural Science Foundation of China (Grants No. 12374135, No. 12374136, No. U1932217, No. 12204487), and the Natural Science Foundation of Shandong Province (No. ZR2022QA040, No. ZR2023QA057).

W.W and Y.X. contributed equally to this work.

W.W. and Y.X. designed and wrote the manuscript with the help of Q.H.,Y.S. and Z.S. All authors have read and agreed to the published version of the manuscript.

\textbf{Conflicts of Interest:}  The authors declare no conflict of interest.

\bibliographystyle{unsrt}%
\bibliography{ref}

\end{document}